\documentclass[12pt]{article}
\usepackage[a4paper, total={6.8in, 10in}]{geometry}
\usepackage[cmex10]{amsmath}
\usepackage[pdftex]{graphicx}
\usepackage{natbib}
\usepackage{hyperref}
\usepackage{parskip}
\usepackage{flafter}

\hypersetup{pdfinfo={
	Title={A method for Bayesian regression modelling of composition data},
	Subject={Bayesian regression modelling for composition data},
	Author={Sean van der Merwe},
	Keywords={Compositional data, proportions, Dirichlet distribution, regression, Bayes, simulation} } }

\newlength{\figwidth} 

\newcommand{\bx}{\ensuremath{\mathbf{x}}}
\newcommand{\by}{\ensuremath{\mathbf{y}}}

\usepackage{authblk}
\title{A method for Bayesian regression modelling of composition data}
\author[a,*]{Sean~van~der~Merwe}
\affil[a]{University of the Free State, Box 339, Bloemfontein, 9300, South Africa}
\affil[*]{Corresponding author --- vandermerwes@ufs.ac.za}

\begin{document}

\maketitle
\bibliographystyle{agsm}

\begin{abstract}
Many scientific and industrial processes produce data that is best analysed as vectors of relative values, often called compositions or proportions. The Dirichlet distribution is a natural distribution to use for composition or proportion data. It has the advantage of a low number of parameters, making it the parsimonious choice in many cases. In this paper we consider the case where the outcome of a process is Dirichlet, dependent on one or more explanatory variables in a regression setting. We explore some existing approaches to this problem, and then introduce a new simulation approach to fitting such models, based on the Bayesian framework. We illustrate the advantages of the new approach through simulated examples and an application in sport science. These advantages include: increased accuracy of fit, increased power for inference, and the ability to introduce random effects without additional complexity in the analysis.
\end{abstract}

\noindent%
{\it Keywords:}  Compositional data, proportions, Dirichlet distribution, regression, Bayes, simulation

\section{Introduction} \label{secIntro}

When vectors are measured in whole numbers, say based on a classification process, analysis is often based on the Multinomial distribution, with the total count being seen as a nuisance parameter. In cases where the total count is not relevant it seems more natural to work directly with the observed proportions. Sometimes the proportions themselves are observed directly, rather than counts and totals. Any situation where the quantity surveyed or analysed does not affect the expected vector (but may affect the precision) falls in this class of problem.
 
The Dirichlet distribution has been widely accepted in literature for modelling compositional data, subject to the constraint that all the correlations between variables are negative \citep{Maier2014}. A wider class of distributions which allows for positive correlations defined on the same sample space is the Logistic-Normal (LN) \citep{Aitchison1980}. However, the LN distribution has many parameters to estimate due to the unknown covariance parameter matrix. In contrast, the Dirichlet distribution has only $P$ unknown parameters to estimate for $P$ compositional variables. Further, when the model is expanded to the regression framework then the requirement of negative correlations no longer applies \citep{Maier2014}. 

The only remaining restriction of relevance is that observed values should be strictly positive. The occasional zero will not affect results, but the method cannot handle binary data in the dependent variable vectors. No restrictions apply to the explanatory variables --- they may have any structure.

As examples of applications, consider the work of \citet{boukal2015analyses}, who look at the development of insects, or the work of \citet{EspinGarcia2014} who look at genetic analysis problems, or the work of \citet{smithson2006better} who discuss applications in psychology. For an industrial application see \citet{DEWAAL2016} where the composition of coal is analysed. Another area in which Dirichlet regression may be useful is in politics, where only the proportion of voters supporting a set of candidates is of interest, possibly dependent on district or demographics. In ecology, preferences of animals for specific types of prey is of interest. To generalise, any situation where people choose between a set of options, and the researcher is interested in the choice and not the number of people making a choice, then the Dirichlet regression model may be of use.

The paper is outlined as follows: In \autoref{secDirichlet}, a brief introduction on applicable Dirichlet properties is given. After that (\autoref{secDirReg}) we consider existing approaches to the problem of regression with dependent variables that follow the Dirichlet distribution (conditional on explanatory data). Then we introduce the new methodology in \autoref{newmethod}, and illustrate its usefulness via simulated examples (\autoref{sims}). We then analyse movement data arising from a school netball tournament as an example based on observed data (\autoref{secRealdata}). Conclusions and future work are discussed in \autoref{secConclusion}.

\section{The Dirichlet distribution} \label{secDirichlet}

If $\mathbf{Y}$ is distributed Dirichlet$(\alpha_1,\dots,\alpha_{P})$, denoted by $D(\boldsymbol\alpha)$, then the joint density is given by 
\begin{equation} \label{eq:D1}
f(\mathbf{y})= \frac{\prod_{i=1}^{P}\Gamma(\alpha_i)}{\Gamma(\alpha_0)}\left\{ \prod_{i=1}^P y_i^{\alpha_i-1}\right\}\ ,\ 0 < y_i < 1,\ \sum_{i=1}^P y_i = 1,\ \alpha_0 = \sum_{i=1}^{P} \alpha_i 
\end{equation}
\citet{Aitchison1986} calls this distribution the Compositional Dirichlet defined on the specified simplex.

\citet{Wilks1962} and \citet{deGroot1970} provide detailed discussions on many of the properties of the Dirichlet distribution. We mention some relevant properties here.

\begin{enumerate}
\item The means and covariances are
\begin{equation} \label{eq:cov}
\begin{aligned}
E[Y_i]&= \mu_i=\frac{\alpha_i}{\alpha_0}, i=1,\ldots,P \\
\sigma_{ij}&=\left\{\begin{array}{l}
\dfrac{-\alpha_i\alpha_j}{\alpha_0^2(1+ \alpha_0)}, i\neq j\vspace{1ex} \\
\dfrac{\alpha_i(\alpha_0 - \alpha_i)}{\alpha_0^2(1+ \alpha_0)}, i=j
\end{array}\right.
\end{aligned}
\end{equation}
 
$\boldsymbol\mu = (\mu_1,\dots,\mu_p)$ denotes the mean of the distribution and $\Sigma=(\sigma_{ij})\ i,j=1,\dots,P$ the covariance matrix.

\item The marginal distribution of any subset $\mathbf{W}$ of $\mathbf{Y}$ is again Dirichlet$(\alpha_{\mathbf{W}}, \alpha_0 - \sum_{i\in {\mathbf{W}}} \alpha_i)$ \citep{Aitchison1986}.

\item In \autoref{eq:cov}, if $\boldsymbol\alpha$ is a multiple of $\boldsymbol\beta$ for two Dirichlet distributions $D(\boldsymbol\alpha)$ and $D(\boldsymbol\beta)$, the means are the same, but the covariance matrices differ.

\item Fitting the Dirichlet distribution to a sample of reasonable size can be done by following the method of moments or method of maximum likelihood (ML) as described in \citet{Minka}, or by using the Bayesian approach \citep{VanDerMerwe2013}. Note that the ML approach is iterative, while the Bayesian approach uses simulation. Both methods are fast in most cases due to the neat convex nature of the likelihood.

\end{enumerate}

\section{Dirichlet Regression} \label{secDirReg}

In \citet{Carmargo2012} they define the problem as follows:

Let $Y=(\by_{1\cdot };\by_{2\cdot};\dots ;\by_{n\cdot})$ be a sample of vector observations. They use the $i\cdot$ notation to indicate that the vectors are arranged in rows of the matrix $Y$, for practical convenience. Let $X=(\bx_{1\cdot};\bx_{2\cdot};\dots ;\bx_{n\cdot})$ be $Q$ explanatory variables arranged the same way (each column of $X$ is a variable and each row of $X$ corresponding with the same row of $Y$). Recall that $\sum_{j=1}^P y_{ij}=1,\ y_{ij}>0$; while the values of $X$ could be any real numbers. We will use their notation going forward.

Based on the work of \citet{campbell1987multivariate}, \citet{Hijazi2009} and \citet{Carmargo2012} we model each parameter as a linear function of the explanatory variables. In terms of a single observation $\by_{i\cdot}=(y_{i1},\dots,y_{iP})\sim D(\alpha_{i1},\dots,\alpha_{iP})$,
\begin{equation} \label{eq:basicDirReg}
\alpha_{ij}=x_{i1}\beta_{1j}+\dots+x_{iQ}\beta_{Qj}=\bx_{i\cdot}\boldsymbol\beta_{\cdot j}
\end{equation}

Thus, the parameters to be estimated are all the $\beta_{kj},\ k=1\dots Q,j=1\dots P$, subject to the constraint $\alpha_{ij}>0\ \forall\ i=1\dots n,j=1\dots P$. They describe a custom optimisation procedure to estimate these parameters under these constraints. Finally, \citet{Carmargo2012} explain an approach to testing $\beta_{kj}=0$, which is useful in many problems.

\citet{gueorguieva2008} propose using a log link in each dimension, thus eliminating constraints in the optimisation procedure.

However, each $\beta_{kj}$ does not have a clean interpretation in the above models as $E[Y_{ij}]$ is a function of all $\beta_{kj}$. The difficulty of interpretation is seen as a major drawback by many researchers, and led to the investigation of alternatives.

\citet{Maier2014} applies a multivariate transformation to the parameters of the Dirichlet distribution, arriving at an alternative formulation that has the advantage of modelling the expected value of an observation separately from its precision.

He begins by defining a parameter $\phi=\alpha_0$ to denote the precision of an observation. Looking at \autoref{eq:cov} we see that it is not exactly the precision but acts like the precision in the sense that, for large values, if the value of $\phi$ doubles while the mean vector is unchanged then the variance halves. We note the relationship $\boldsymbol\alpha=\boldsymbol\mu \phi$. \citet{Maier2014} applies a log link function to model $\phi$, \textit{i.e.}\ $\ln \phi_i = \mathbf{w}_{i\cdot}\boldsymbol\beta_{\cdot\phi}$, where $W$ is a matrix of explanatory variables for the precision.

For the purpose of modelling the mean \citet{Maier2014} uses a multivariate logit link. This involves choosing a base category and setting all coefficients to zero for this category, then using a log link to model the other categories and rescaling the results so that the means sum to one. In the notation defined previously, and using dimension 1 as the base, we have that
\begin{equation} \label{eq:multilogit}
\begin{aligned}
\mu_{ij}&=\frac{\exp (\bx_{i\cdot}\boldsymbol\beta_{\cdot j})}{\sum_{k=1}^P \exp (\bx_{i\cdot}\boldsymbol\beta_{\cdot k})} \ ,\ j=2\dots P \\
\mu_{i1}&=\frac{\exp \bx_{i\cdot}\mathbf{0}}{\sum_{k=1}^P \exp \bx_{i\cdot}\boldsymbol\beta_{\cdot k}} = \frac{1}{1+\sum_{k=2}^P \exp \bx_{i\cdot}\boldsymbol\beta_{\cdot k}}
\end{aligned}
\end{equation}

\citet{Maier2014} explains that using the transformation above results in coefficients that are interpretable as odds ratios if exponentiated.

However, each $\mu_{ij}$ is still a function of all $\beta_{kj}$, and we don't have coefficients for the base dimension. These limitations inhibit interpretation. In the next section we introduce a new methodology that incorporates the best aspects of the approaches described above.

\section{New methodology} \label{newmethod}

The first change is the use of a univariate logit transformation for each mean parameter individually, thus allowing all $\beta_{kj}$ coefficients to be unrestricted real numbers. The second change that ties in with the first is that we abandon the idea of a reference category or dimension.

In theory, one dimension is redundant since it is a linear combination of the others, but in practice we desire to know the relationship between the explanatory variables and the outcomes in all dimensions. Often the dimensions are equal in the view of the researcher and it is not sensible to relegate one to reference status. It is for this reason that researchers such as \citet{chen2016} resort to modelling each dimension individually as Beta distributed random variables, but that in turn ignores the multivariate nature of the data.

It is desirable to have each $\beta_{kj}$ relate directly to a single dimension in a way that can be directly interpreted. By modelling all dimensions we move closer to this ideal.

The third change is a move to the Bayesian framework. We introduce vague Normal priors on all $\beta$ parameters. All other parameters are defined in terms of these, so no further priors are necessary at this stage.

Specifically, we use the Bayesian simulation framework, which holds many advantages. It allows us to directly quantify uncertainty in both the coefficients and the means. Also, when moving to a predictive framework, construction of predictive densities is relatively straightforward.

We are still bound by the conditions 
\begin{equation} \label{eq:restrict}
\sum_{j=1}^P \mu_{ij} = 1\ \forall\ i=1\dots n
\end{equation}
which impede the standard simulation approach greatly. In order to have the simulation process run smoothly, we must introduce a source of flexibility into the model. We choose to add flexibility by replacing the restriction (\autoref{eq:restrict}) by a penalty on the likelihood:
\begin{equation} \label{eq:penalty}
L^{*}\propto L\times \exp\left\{-\frac{1}{\xi}\sum_{i=1}^n \left[\left(\sum_{j=1}^P \mu_{ij}\right) -1\right]^2\right\}
\end{equation}

The hyperparameter $\xi$ must be chosen large enough to allow the simulation procedure to run smoothly, but small enough to have minimal impact on the simulation results. By minimal impact we mean deviations that can easily be corrected. The valid region for $\xi$ to meet these criteria seems surprisingly large. One may consider $\xi$ as a hyperparameter and choose its value manually, or, more conveniently, apply a prior distribution to $\xi$ and have the value vary as part of the simulation process.

Then, for further simulation flexibility we introduce a second penalty parameter $(\xi^{*})$ in the relationship between $\boldsymbol\alpha$, $\boldsymbol\mu$ and $\phi$. For both penalty parameters we found that a simple Exponential prior works well in all scenarios tested.

Assuming explanatory data captured in matrices $X$ and $W$ (which may overlap), we define the model in a hierarchical fashion:
\begin{equation} \label{eq:hierarchy}
\begin{aligned}
\by_{i\cdot}|\bx_{i\cdot},\mathbf{w}_{i\cdot} &\sim D(\boldsymbol\alpha) \\
\ln \alpha_{ij} &\sim N\left(\ln \mu_{ij} + \ln \phi_i\ ,\ \frac{1}{\xi^{*}}\right) \\
\ln \phi_i &= \mathbf{w}_{i\cdot}\boldsymbol\beta_{\cdot\phi} \\
\text{logit}(\mu_{ij}) &= \bx_{i\cdot}\boldsymbol\beta_{\cdot j} \\
\sum_{j=1}^P \mu_{ij} &\sim N\left(1,\frac{1}{\xi} \right) \\
\beta_{ij}\ ,\ \beta_{i\phi} &\sim N(0,10000) \\
\xi^{*} &\sim Exp(\mu=100/P) \\
\xi &\sim Exp(\mu=1000/P) \\
\end{aligned}
\end{equation}
Note that we express the models for the mean $(\text{logit}(\mu_{ij}) = \bx_{i\cdot}\boldsymbol\beta_{\cdot j})$ and precision $(\ln \phi_i = \mathbf{w}_{i\cdot}\boldsymbol\beta_{\cdot\phi})$ in linear form for easy of understanding only --- these models can be extended as needed by the researcher.

We implement the model using Gibbs sampling \citep{Gelfand1990} via the OpenBUGS program (\url{http://www.openbugs.net/}). Implementation is done indirectly through the R2OpenBUGS package \citep{R2OpenBUGS} for R \citep{RCore}. All pre- and post-calculations are done in R. The MASS and parallel packages supplied with R were also used to facilitate calculations.

Post-simulation, we apply the following corrections to each simulated parameter set $k$ individually to ensure that fitted expected values sum to one for each observation:
\begin{equation} \label{eq:corrections}
\begin{aligned}
\mu^{adj}_{ijk} &= \frac{\mu^{sim}_{ijk}}{\sum_{j=1}^P \mu^{sim}_{ijk}} \\
\alpha^{adj}_{ijk} &= \mu^{adj}_{ijk}*\phi^{sim}_{ik}
\end{aligned}
\end{equation}

\section{Simulation Study} \label{sims}

Since the different methods discussed in the previous sections use different transformations (identity, log, multivariate logit, and univariate logit), we compare methods on a single scale. In general the researcher fitting Dirichlet regression models is interested in three things: the significance of the coefficients, the direction of any significant relationships, and the accuracy of the model fit on the observed data. We focus on model accuracy first.

As a measurement of error we consider the average \textit{sum of compositional errors} (SCE), explained by \citet{Hijazi2009}. It is the sum of the Aitchison distances \citep{Aitchison1986} between estimated compositions and the target values.

Further, we calculate the intervals for each expected value individually and then report the average coverage, along with the average width (standardised by dividing by the expected values). We also consider coverage using the posterior predictive distribution.

Datasets are constructed from a model exactly in line with the `alternative' specification of \citet{Maier2014}. Models are then fitted using his DirichletReg package, as well as the new methodology. Models are correctly specified in all cases --- model misspecification is beyond the scope of this work.

\subsection{Simulation Scenario A} \label{sec:scenA}

Scenario A is a simple analogy to the MANOVA problem. Consider a single explanatory variable that is a factor with three levels.  A researcher might be interested in whether the mean vectors differ between the three groups created by the factor levels, under the assumption of constant variance.

Let the observed vectors have three dimensions. So we set the coefficients for the first dimension to zero, and then use coefficients of (-0.9,0.6,1.2) for the first dimension and (0.8,-1,0.5) for the second. The inverse multivariate logit transformation is then applied to obtain the `true' expected values for every observation, and expanded to 20 observations per factor level ($n=60$ in total). The first step is creating a matrix of binary variables $(X)$ from an expansion of the explanatory factor. The transformation (explained in detail in \cite{Maier2014}) involves multiplying $X$ by each set of coefficients, and then exponentiating to obtain raw expected values, which are then standardised to sum to one for each observation.

The next step is to multiply by a chosen value for $\phi$. We used $\phi=1$ for illustration. The effect of changing this value will be discussed after the results. Multiplying the expected values by $\phi$ yields a matrix of $\alpha_{ij}$ values, which is used to simulate hundreds of Dirichlet samples in the standard way.

After every sample is modelled and the results summarised, various statistics are produced. See \autoref{tb:ScenA} for the most important values. It is clear that the new methodology produces better fit.

\begin{table}[htb]\begin{center}
\begin{tabular}{|c|c|c|c|} \hline
Scenario A & Target & Maier approach & New approach  \\ \hline
Error  & 0 & 19.59 & 18.38 \\ \hline
Coverage & 0.95 & 0.87 & 0.94 \\ \hline
Std. Width & 0 & 0.70 & 0.75 \\ \hline
\end{tabular}\end{center}
\caption{Fit statistics for Scenario A} \label{tb:ScenA}
\end{table}

The next question of interest would be the effect of varying the chosen parameter values. It appears that the important parameter is the underlying precision, which is closely related to the $\phi$ parameter. When the data is measured with high precision (say $\phi\geq 5$) then there is little difference between the fits obtained by the two methods (both methods fit very well). As the precision drops and the underlying relationships become more obfuscated then the accuracy of the previous methodology falls away rapidly, while the new methodology loses accuracy slowly, resulting a the difference observed in \autoref{tb:ScenA}.

Another question is interest would be the effect of increasing the dimensionality of the problem. If the dimension is increased to 8 and the coefficients are chosen as random $U(-1.5,1.5)$ values, then the relative differences between the methods become even more exaggerated. The new methodology adapts easily to having many categories in the dependent variable.

\subsection{Simulation Scenario B} \label{sec:scenB}

Scenario B is a more complex scenario where a linear term is introduced in every mean vector as well as the precision model, in addition to the factor explained in \autoref{sec:scenA}. The goal is to determine whether the model can identify both relationships simultaneously in all categories.

The explanatory factor $(X_1)$ is given two levels with coefficients (0,0), (-0.9,0.6), (1.8,-1) in three dimensions respectively. The explanatory variable with a linear relationship $(X_2)$ is given real values between 4.5 and 7.5 with 40 values per level of the explanatory factor. The linear relationship is created by adding $0.75X_2$ to the third dimension and then correcting the means to add to one, thus creating a positive relationship in the third dimension and implicitly creating negative relationships in the first two dimensions. As $X_2$ increases $y_1$ and $y_2$ will tend to decrease, while $y_3$ will tend to increase. The expression for $\log \phi$ used to generate the data is $-1+0.5X_2$.

Again we average the results from hundreds of samples and summarise the results. As indicated in \autoref{tb:ScenB}, the results from the methods are similar, but the new methodology is more accurate. 

\begin{table}[htb]\begin{center}
\begin{tabular}{|c|c|c|c|} \hline
Scenario B & Target & Maier approach & New approach  \\ \hline
Error  & 0 & 19.19 & 18.81 \\ \hline
Coverage & 0.95 & 0.85 & 0.86 \\ \hline
Std. Width & 0 & 0.52 & 0.54 \\ \hline
\end{tabular}\end{center}
\caption{Fit statistics for Scenario B} \label{tb:ScenB}
\end{table}

Concerning inference, the new method shows a major improvement over the existing method. The new method correctly identifies the direction of the linear relationships and marks all of them as significant. The existing method only identifies the linear relationship in the precision model. We base this judgement on the median p-values obtained over the many simulations.

In the precision model the median p-value for the existing method is 0.1\%, while for the new method it is approximately 0\%. In the second and third dimensions the median p-values for the existing method are 50\% and 24\%, while for the new method we obtain 1\% and 0.1\%. The new method also reports a p-value for the first dimension $(0.4\%)$, while the existing method does not.

\section{Observed data example from sport science} \label{secRealdata}

During a school netball tournament, scholars were tracked accurately as they move across the field. One of the resulting measurements was the proportion of time spent standing/walking/running during the course of a match. The goal is to investigate the relationship between these measurements and the playing position. 

A major complicating factor is the fitness and behaviour variation between players. Some players were observed for only one match, while others were observed for up to nine matches. This suggests an unbalanced mixed effects model, with position as fixed effect and player as random effect.

We begin by implementing simple approaches that might be used to analyse this data. This is to illustrate the practical differences between approaches. First, we fit linear models on the logit scale to each dimension separately. This is essentially the same as using descriptive statistics to analyse the data. It fails to account for the multivariate nature of the observations and for the player effect. Second, we fit mixed effects models in each dimension (on the logit scale) to account for the player effect. The mixed effects model fitted did not account for the multivariate nature of the data.

Third, we implement the method of \citet{Maier2014}. This accounts for the multivariate structure, but not the player effect; and then finally we implement the new approach, which takes into account both the structure of the data and the player effect at once.

\begin{figure}
\centering
\includegraphics[width=0.9\textwidth]{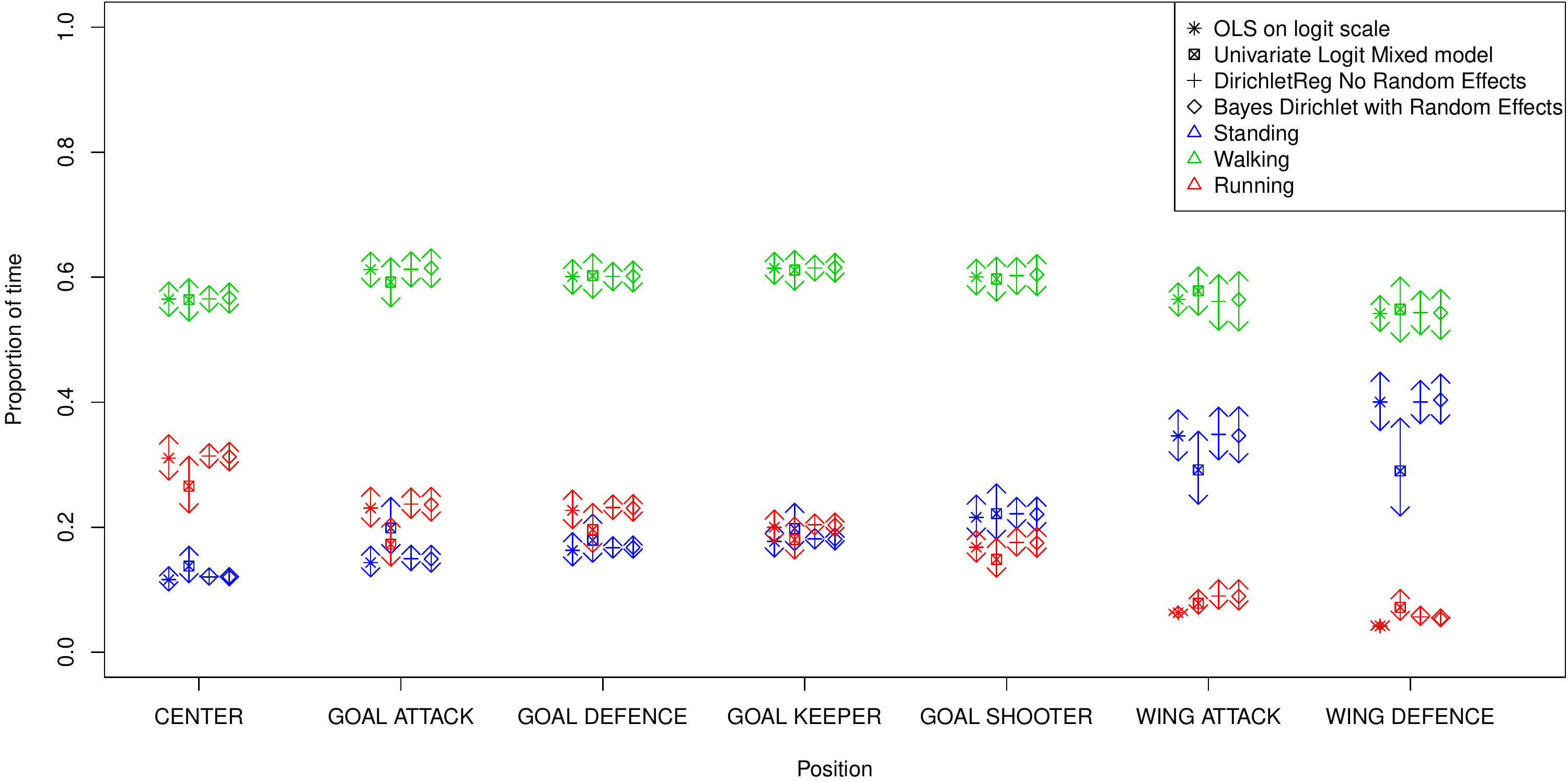}
\caption{Differences in movement between playing positions in a netball tournament as suggested by different modelling approaches.} \label{fig:netball}
\end{figure}

In \autoref{fig:netball} we see the different methods discussed and the differences between them. The results produced by the new methodology appear sensible in both the expected values and intervals.

\section{Conclusions} \label{secConclusion}

In this paper the goal was a generic solution to the problem where the outcome of a process is Dirichlet, dependent on one or more explanatory variables in a regression setting. Existing approaches were discussed and a new methodology introduced. The new methodology was directly compared to the latest of the existing approaches and found to perform well. At worst the performance is in line with existing tools, but in many cases the improvement is remarkable, especially when the data has high variance. Advantages of the new methodology were discussed, including ease of interpretation and prediction, with accurate intervals, as well as the ability to introduce random effects.

\section*{Acknowledgements}

The netball data is from a study by Michael Shaw and Derik Coetzee of the Department of Exercise and Sports Science, University of the Free State, South Africa. The author wishes to thank Robert Schall and Michael von Maltitz for useful advice during the development of this work.

\bibliography{DirichletRegression}

\end{document}